%% file: main.tex
\documentclass[showpacs,prl,aps,10pt,superscriptaddress,altaffilletter,floatfix,twocolumn]{revtex4-1}
\usepackage{graphicx}
\usepackage{hyperref}
\usepackage{amsmath}
\usepackage{amssymb}
\usepackage{color}
\usepackage{xspace}
\usepackage{dcolumn}
\usepackage{soul}
\usepackage{comment}
\newcolumntype{C}[1]{>{\centering\let\newline\\\arraybackslash\hspace{0pt}}m{#1}}
\usepackage[section]{placeins}

\RequirePackage{lineno}

\usepackage{upgreek}

\begin{document}

\input{commands}
\setstcolor{blue}

\title{MoEDAL search in the CMS beam pipe for magnetic monopoles produced via the Schwinger effect}

\include{moedal_authors_revtex}

\date{\today}

\begin{abstract}
We report on a search for magnetic monopoles (MMs) produced in ultraperipheral Pb--Pb collisions during Run-1 of the LHC. The beam pipe surrounding the interaction region of the CMS experiment was exposed to 184.07 \textmu b$^{-1}$ of Pb--Pb collisions at 2.76 TeV center-of-mass energy per collision in December 2011, before being removed in 2013. It was scanned by the MoEDAL experiment using a SQUID magnetometer to search for trapped MMs. No MM signal was observed. The two distinctive features of this search are the use of a trapping volume very close to the collision point and ultra-high magnetic fields generated during the heavy-ion run that could produce MMs via the Schwinger effect. These two advantages allowed setting the first reliable, world-leading mass limits on MMs with high magnetic charge. In particular, the established limits are the strongest available in the range between 2 and 45 Dirac units, excluding MMs with masses of up to 80 GeV at 95\% confidence level.
\end{abstract}


\maketitle

The symmetry of the Maxwell’s equations under exchange of the electric and magnetic fields, currently apparent only in vacuum, would be restored if isolated magnetic charges, or magnetic monopoles (MMs), exist. The presence of MMs could explain the quantization of the electric charge, as shown by Dirac~\cite{Dirac:1931}. Dirac's argument also requires the magnetic charge to be quantized, with the fundamental unit of $g_{\mathrm{D}}$ = 2$\pi\hbar/e$, the Dirac charge. Further motivation for MMs is provided by a number of Grand Unified Theories (GUTs)~\cite{Hooft:74,Polyakov:1974ek} and other beyond the standard model (BSM) scenarios~\cite{CHO:1997, cho:2015, Ellis:2016, mavromatos:2017, Arunasalam:2017, Mavromatos:2018, Hung:2020, Shafi:2022} that contain topologically stable, finite-energy MM solutions, some with expected masses on the order of a TeV. Importantly, in such models the fundamental magnetic charge is predicted to be two or three times larger than the Dirac charge, the latter possibility realized if the MM does not carry magnetic color charge~\cite{Shafi:2021}. In general, the value of the minimum magnetic charge depends on the global properties of the gauge group of the underlying theory~\cite{Tong:2017oea}, with an $SU(N)$ theory typically giving rise to a minimal magnetic charge of $N$ Dirac units. Finding the minimum magnetic charge could provide a unique way to probe the structure of the gauge group.
Finally, MMs predicted in string theories are also required to have a fundamental magnetic charge equal to an integer multiple of the Dirac value, as shown by Wen and Witten~\cite{Wen-Witten:1985}.

The motivation for the existence of MMs has led to an array of direct searches utilizing a wide range of techniques and tuned to very different MM velocity and mass ranges (for a recent review, see Ref.~\cite{Mavromatos:2020gwk}). However, many of the searches were limited to smaller magnetic charges, as the ionization losses of MMs in matter increase rapidly with magnetic charge, leading to loss of detection efficiency owing to MM absorption in passive materials or electronics saturation effects. One way to increase sensitivity to high magnetic charges is to use an accelerator beam pipe as an MM trapping volume~\cite{DeRoeck2012,DeRoeck:2012wua}, an approach that allowed the H1, D0, and CDF collaborations to establish limits on MMs with up to six units of Dirac charge~\cite{hera_beampipe:2005, d0_cdf_beampipe:2004, d0_beampipe:2000}. Another exception to the limitation is the MoEDAL experiment, whose dedicated sensitivity to high ionization allowed it to set limits on particles with magnetic charge of up to five $g_{\mathrm{D}}$~\cite{moedal_prl2019,moedal_dyons}. The most recent MoEDAL result
has extended the sensitivity to ten units of Dirac charge~\cite{HECO2_arxiv}.

An additional difficulty of most collider experiments to date is establishing reliable mass limits on MM production. Because of the large coupling of MMs to photons, perturbation theory cannot be used to calculate the production cross section of MMs in elementary particle collisions (e.g., parton-parton collisions that are typically studied at hadron colliders) unless appropriate resummation schemes are applied~\cite{PhysRevD.100.096005}. 
It is also expected that the production of composite MMs --- the type predicted by the GUT and BSM models --- is exponentially suppressed by a factor of e$^{-{\cal O}(500)}$ in such collisions~\cite{Witten:1979,drukier}. This is because the physical size of a composite MM is larger than its Compton wavelength by a factor of $(4\pi)/e^2$~\cite{goebel1970spatial, Goldhaber:1982qh}, and hence an ultrarelativistic particle collision above the energy threshold for MM production is localized to a region much smaller than the MM size. The matrix elements for production are weighted by the wave functions of the initial and final states, and these have exponentially small overlap. These two limitations were recently overcome by a search for MM production in collisions of lead ions at the LHC via the Schwinger effect~\cite{MoEDAL:2021vix}. 

The Schwinger effect is a semi-classical process that describes the generation of particle-antiparticle pairs in the presence of a strong electromagnetic field. In particular, the strong magnetic fields generated during ion collisions may decay via quantum tunneling to form a magnetic monopole-antimonopole pair~\cite{Gould:2021bre}. The production cross section depends strongly on the amplitude of the magnetic field. In heavy-ion collisions, the field reaches its maximum at large impact parameters, when the centers of the ions are separated by almost twice the nuclei radius. Such collisions are called ultraperipheral~\cite{UPC}. Importantly, the Schwinger production cross section is calculable nonperturbatively, and the production of composite MMs is actually enhanced compared to elementary MMs, thanks to the coherence length of the magnetic field in UPCs being larger than the MM size~\cite{Gould:2021bre,arttu_prd2019}. While proton-proton collisions may generate even stronger fields thanks to their higher Lorentz factors, the total volume and energy of the generated field are small, which limits the production of MMs. 

The search in Ref.~\cite{MoEDAL:2021vix} established limits for MMs with up to three units of Dirac charge, its sensitivity to higher charges being hampered by the absorption of the highly-ionizing MMs by passive materials before they could reach the detectors. In order to extend greatly the sensitivity to high magnetic charge while still being able to calculate reliably the production cross section, we search for MMs trapped in the CMS beam pipe during heavy-ion collisions. The beam pipe was exposed to Pb--Pb collisions at the center-of-mass energy per nucleon-nucleon collision, $\sqrt{S_{\mathrm{NN}}}$, of 2.76 TeV with an integrated luminosity delivered to CMS, $L_\mathrm{CMS}$, of 184.07 \textmu b$^{-1}$ during Run-1 of the LHC. The beryllium pipe is 1 mm thick and extends 1902 mm in both directions from the interaction point. The CMS solenoid provided a near-uniform 3.7-3.8 T magnetic field at the location of the pipe~\cite{CMS_Bfield_2010}. 

The fleeting magnetic field generated during the ultraperipheral collisions is calculated following the approach described in Ref.~\cite{arttu_prd2019}. The field is characterized by its peak strength, $B_\mathrm{max}$, and inverse decay time, $\omega$, estimated, in natural units:
\begin{align}
\label{eqn:field_decaytime}
B_\mathrm{max} &\approx c_{B} \frac{Ze\beta\gamma}{2\pi R_\text{Pb}^{2}},
&
\omega &\approx c_{\omega} \frac{\beta\gamma}{R_\text{Pb}} \, ,
\end{align}
where $R_\text{Pb}$ 
is the lead ion radius, $Ze$ is the lead ion charge, $\beta\approx$1 is the ion velocity as a fraction of the speed of light in vacuum, $\gamma$ is the corresponding Lorentz factor, $c_{B}$ and $c_{\omega}$ are $O(1)$ numerical coefficients. The peak field strength occurs at an impact parameter $b = \textit{b}_\text{max}$  $\approx$ 1.94$R_\text{Pb}$~\cite{arttu_prd2019}.

The production cross section and center-of-mass kinematics are calculated following the formalism developed in Refs.~\cite{arttu_prd2019, Gould:2021bre}. These references describe two approximate approaches --- the free particle approximation (FPA) and the locally constant field approximation (LCFA). 
The FPA approximation considers the spacetime dependence of the electromagnetic fields generated during a collision to all orders but ignores the MM pair's self-interactions. Conversely, the LCFA approximation considers self-interactions to all orders but neglects the spacetime dependence of the electromagnetic field. Both approaches provide conservative lower limits on the production cross section, as including the self-interactions and fields' spacetime dependence has been shown to increase the MM production~\cite{arttu_prd2019, Gould:2021bre}. For high magnetic charges, the LCFA is expected to be more reliable, as self-interactions become more important, and because the spacetime dependence becomes less important at higher charges~\cite{arttu_prd2019}. Consequently, this work uses the LCFA cross section shown in Eq.~(\ref{eqn:LCFA}) as a function of the MM mass, $M$, and magnetic charge, $g$: 
\begin{equation}
\label{eqn:LCFA}
\sigma_{\rm LCFA} = \theta(F-2M) \frac{2(g B_\mathrm{max})^{9/2}R_\text{Pb}^4}{9 \pi^2 M^5 \omega^2}\exp\left(-\frac{\pi M^2}{g B_\mathrm{max}}+\frac{g^2}{4}\right) \, ,
\end{equation}
where $\theta$ denotes the step function and $F$ is the total energy in the electromagnetic field available for MM pair production. This total cross section is integrated over the impact parameter, with the dominant contribution coming from $b = b_\mathrm{max}$, i.e., from the ultraperipheral collisions~\cite{MoEDAL:2021vix}. The total available energy shown in Eq.~(\ref{eqn:maxField}) provides an upper limit on the mass of an MM that could be produced in a collision, regardless of the value of the cross section:
\begin{equation}
\label{eqn:maxField}
  F = \int \, d^{3}x \frac{1}{2} (B^{2}-E^{2}) \theta(B^{2}-E^{2}) \Big|_{t=0} \, ,
\end{equation}
where $B$ and $E$ are the magnetic and electric fields generated during the collision, and the integral is evaluated over the collision region at the time of the closest approach.
A more complete calculation of the cross section in Eq.~(\ref{eqn:LCFA}) would include the backreaction effects of MM pair production on the electromagnetic fields. However, since it includes the negative contribution of the energy of the electric field, $F$ provides a conservative estimate of the mass threshold.

The kinematics of MM production in the Pb--Pb collisions is dominated by the time dependence of the electromagnetic fields. While the LCFA is expected to describe better the total cross section, it neglects the time dependence of the field and leads to an underestimate of the width of the momentum distribution that is incompatible with the time--energy uncertainty principle. Hence, following Ref.~\cite{Gould:2021bre}, we use the FPA formalism to compute the momentum distribution, which is an important input to the simulation used to calculate the probability that a produced MM is captured in the beam pipe.

A detailed \textsc{Geant4}~\cite{geant4} Monte Carlo (MC) simulation is used to estimate the fraction of produced MMs that are captured by the beam pipe. The MMs are generated at the CMS interaction point with the theoretical momentum distribution and propagated through the beam pipe geometry, taking into account the effect of the CMS solenoidal magnetic field and energy losses. The MM's electron ionization energy loss is implemented according to the formalism described in Refs.~\cite{Ahlen:1978, bologna_2016, Ahlen1982}. This provides an accurate description of total energy loss from the relativistic regime down to $\beta \sim$ 10$^{-3}$. The nuclear stopping power, which increases the total loss by $\sim$1\% at $\beta$=10$^{-2}$, has been included in this work following Ref.~\cite{Ahlen1982}. For diamagnetic materials like beryllium, another contribution to the energy loss becomes relevant below $\beta$ $\sim$10$^{-3}$~\cite{DERKAOUI1998173} and was implemented here for MMs with magnetic charges $\leq$2 $g_{\mathrm{D}}$. The nuclear and diamagnetic components of the total energy loss were found to have insignificant effects on the results.

\begin{figure}[htpb]
\centering
\includegraphics[width=0.49\textwidth]{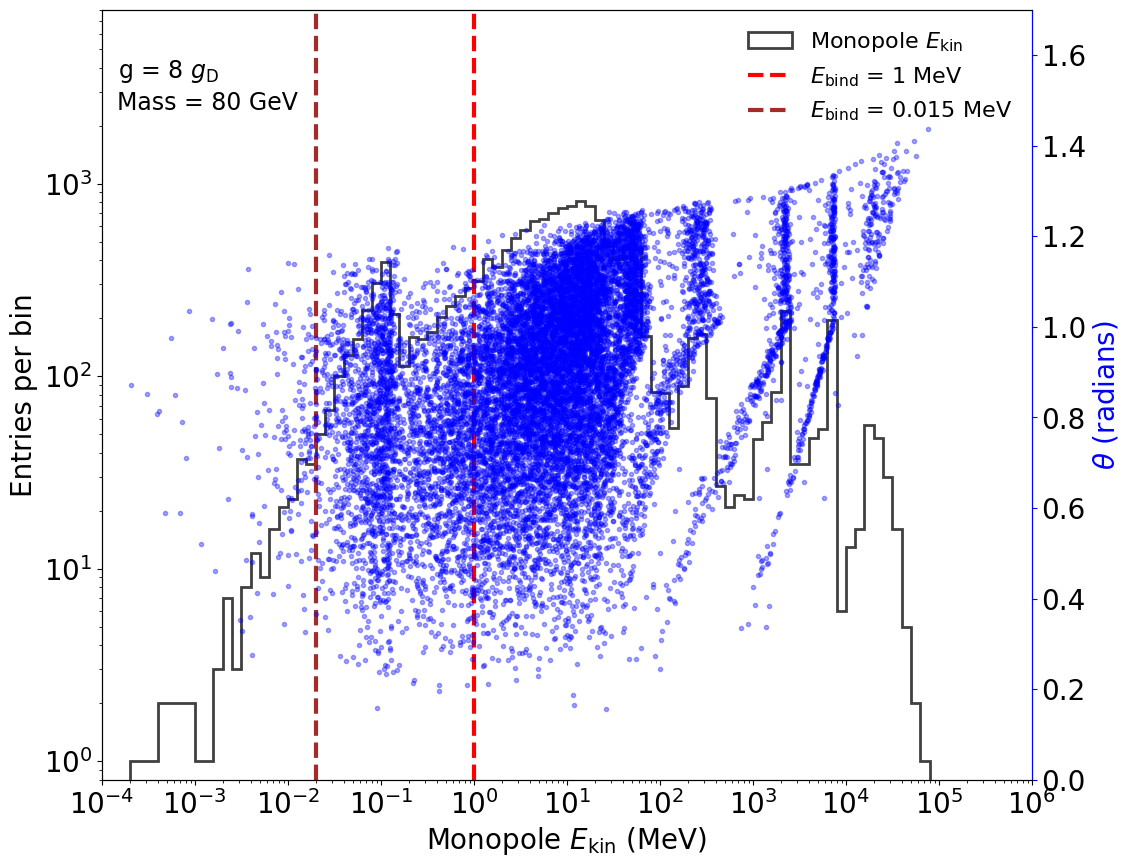}
\caption{The distribution of the MMs' kinetic energies ($E_{\mathrm{kin}}$) within the beam pipe at the point of turning (black histogram). The dashed red and maroon lines show the two assumed binding energies ($E_{\mathrm{bind}}$). The distribution of initial polar angles for MMs that turn inside the beam pipe is also shown (blue dots). The distributions correspond to MMs with 8 $g_\mathrm{D}$ charge and 80 GeV mass.}
\label{fig:BPmomentum}
\end{figure}

Once inside the beam pipe, the MMs begin to lose energy. However, the energy losses in beryllium are insufficient to counteract the acceleration by the CMS magnetic field. Therefore, the only MMs (or anti-MMs) that could potentially get trapped in the beam pipe were those produced with an initial momentum directed against (or along) the field lines. Such MMs are decelerated by the field and eventually turn around toward the field's direction. If, at the instant of turning, the MM's position is within the beam pipe volume, then it would reach its lowest kinetic energy, $E_\mathrm{kin}$, inside the beam pipe material. If this kinetic energy is less than the expected binding energy, $E_\mathrm{bind}$, then the MM is expected to be trapped. According to Ref.~\cite{Milton_2006}, an MM would disrupt the beryllium nucleus and bind to its constituents. Among the different thresholds on binding energy --- binding to an individual proton or neutron, or all nine nucleons comprising the nucleus --- the conservative scenario is binding to a single proton, with a calculated binding energy of 15.1 keV~\cite{BRACCI1984236}, increasing to 1 MeV if the proton's form factor is taken into account~\cite{OLAUSSEN1985465}. Even for the conservative case of 15.1 keV binding energy, the lifetime~\cite{Milton_2006} of an MM--proton bound state in the presence of the CMS external field is estimated to be very large for all values of the magnetic charge considered in this work. Figure~\ref{fig:BPmomentum} shows an example distribution of the MMs' kinetic energies at the instant of turning within the beam pipe for a MM with 8 $g_\mathrm{D}$ charge and 80 GeV mass. The peaks in the distribution coincide with the initial angle $\theta$ relative to the beam direction that are favored by the MMs turning within the beam pipe region. Only specific initial directions of MMs with specific initial kinetic energies will result in them turning inside the beam pipe, giving rise to the features of the distribution shown in the figure.

\begin{figure}[htpb]
\centering
\includegraphics[width=0.49\textwidth]{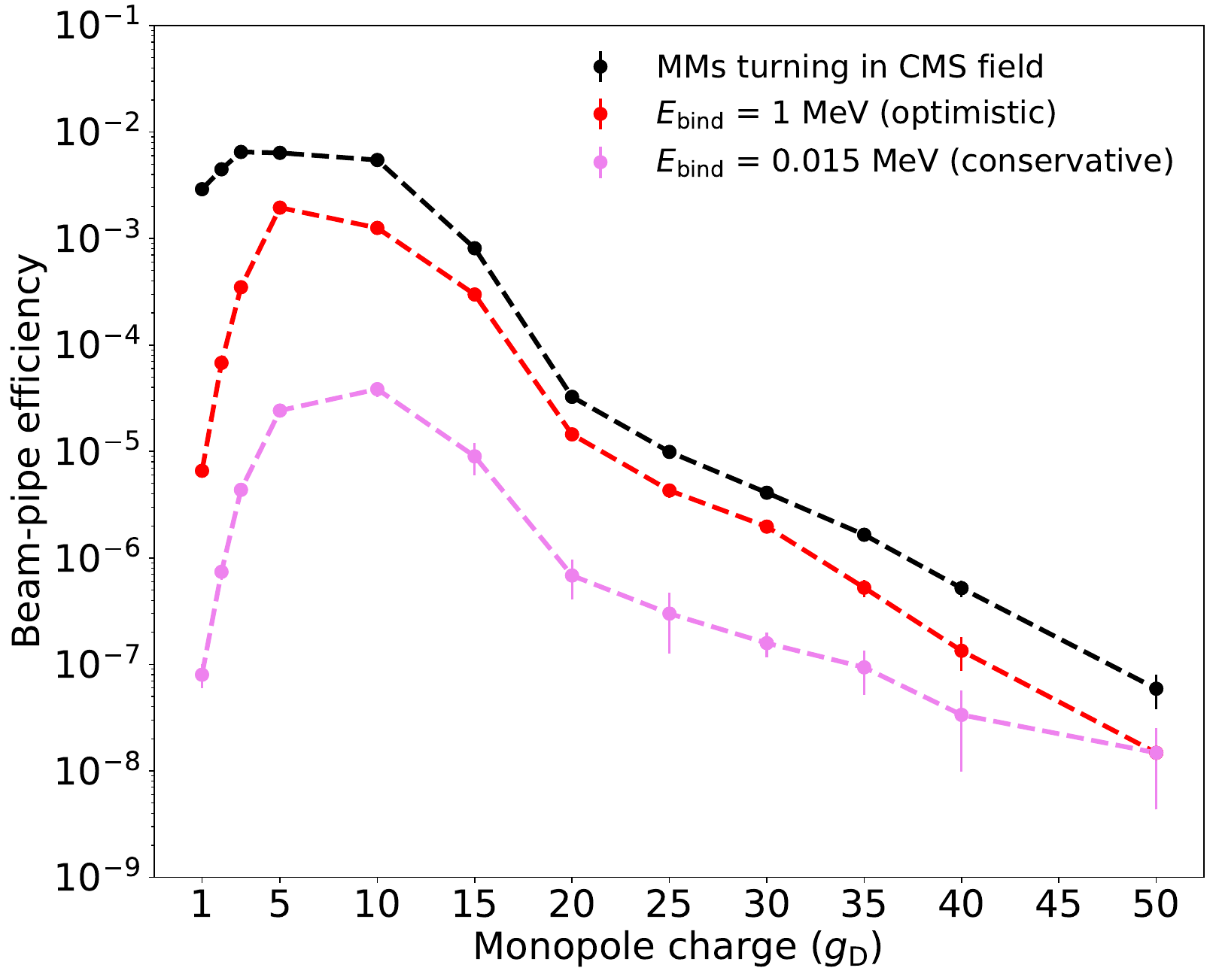}
\caption{The MM trapping efficiency in the CMS beam pipe as a function of the magnetic charge.}
\label{fig:BPefficiency}
\end{figure}

The beam pipe trapping efficiency is calculated as the ratio of the number of MMs bound to the beryllium volume to the total number of generated MMs. The trapping efficiency decreases rapidly for MMs with magnetic charges below 3 $g_{\mathrm{D}}$, because the MMs do not lose enough energy in the 1-mm thick beam pipe and punch through. On the other hand, the efficiency also drops quickly for very high magnetic charges because they would decelerate and turn before reaching the beam pipe, since their acceleration is proportional to the magnetic charge. Thus, the trapping efficiency shown in Fig.~\ref{fig:BPefficiency} peaks at magnetic charges of 5--10 $g_{\mathrm{D}}$. 

We define $R_{\mathrm{exp}}$ as the mean rate of MMs expected to be trapped in the beam pipe during the LHC Run-1 Pb--Pb collisions. For a particular MM mass and charge, the expected rate is calculated as a product of the trapping efficiency, the CMS integrated luminosity, and the LCFA cross section. Two bounds on the MM--nucleus binding energy --- 15.1 keV and 1.0 MeV --- are used in the calculation as described earlier. Overall, the results are not very sensitive to this choice because the expected rate for charges above 2 $g_{\mathrm{D}}$ remains high enough that MMs with the mass corresponding to the upper limit in Eq.~(\ref{eqn:maxField}) could still be excluded in spite of the lower trapping efficiency. 

The main systematic error in $R_{\mathrm{exp}}$ is due to the uncertainties in the peak magnetic field strength $B_\mathrm{max}$ and the inverse decay time $\omega$. For Pb--Pb collisions at $\sqrt{S_{NN}}$=2.76 TeV, the peak value of $B_\mathrm{max}$ = 4.0 $\pm$ 0.2 GeV$^{2}$ and $\omega$ = 40 $\pm$ 1.7 GeV is calculated using Eq.~(\ref{eqn:field_decaytime}). These uncertainties, described in detail in Ref.\cite{MoEDAL:2021vix}, arise from the assumptions of the fit model and are reflected in the uncertainty of the results. We assume uniform distributions within these ranges of uncertainty. The final positions of MMs with identical initial momenta were checked to be stable to within 1\% of the beam pipe thickness when the \textsc{Geant4}'s step length and field integration parameters ($\epsilon_{max}$, ``delta intersection'', ``delta one step'', and ``deltachord'') were varied within recommended bounds~\cite{geant4guide}. The systematic uncertainty in the MM energy loss calculated using \textsc{Geant4} is $\approx$10\%~\cite{MoEDAL2016}, which leads to a negligible variation in trapping efficiency compared to the effect of the different binding energy assumptions.

\paragraph{Results and conclusion.}

The CMS beam pipe exposed to the Pb--Pb collisions during Run-1 of the LHC was scanned for the presence of trapped MMs using a direct current SQUID long-core magnetometer installed at ETH Zurich. The north and south poles of a magnetic dipole passing through the magnetometer would induce currents that cancel each other out. In contrast, the presence of an MM in the sample would induce a persistent current in the superconducting coil, directly proportional to the strength of the magnetic pole. This effect does not depend on the mass of the MM, only on the magnetic charge. Several instrumental and environmental effects could lead to spurious readings, potentially mimicking an MM signal. These could be caused by physical vibrations and shocks, variations in external magnetic fields, passage of a large sample through the sensing coil at a high speed, and other effects~\cite{MoEDAL2016}. Repeated measurements on the sample would average out such spurious signals, while a MM would consistently yield the same value.

\begin{figure}[htpb]
\centering
\includegraphics[width=0.49\textwidth]{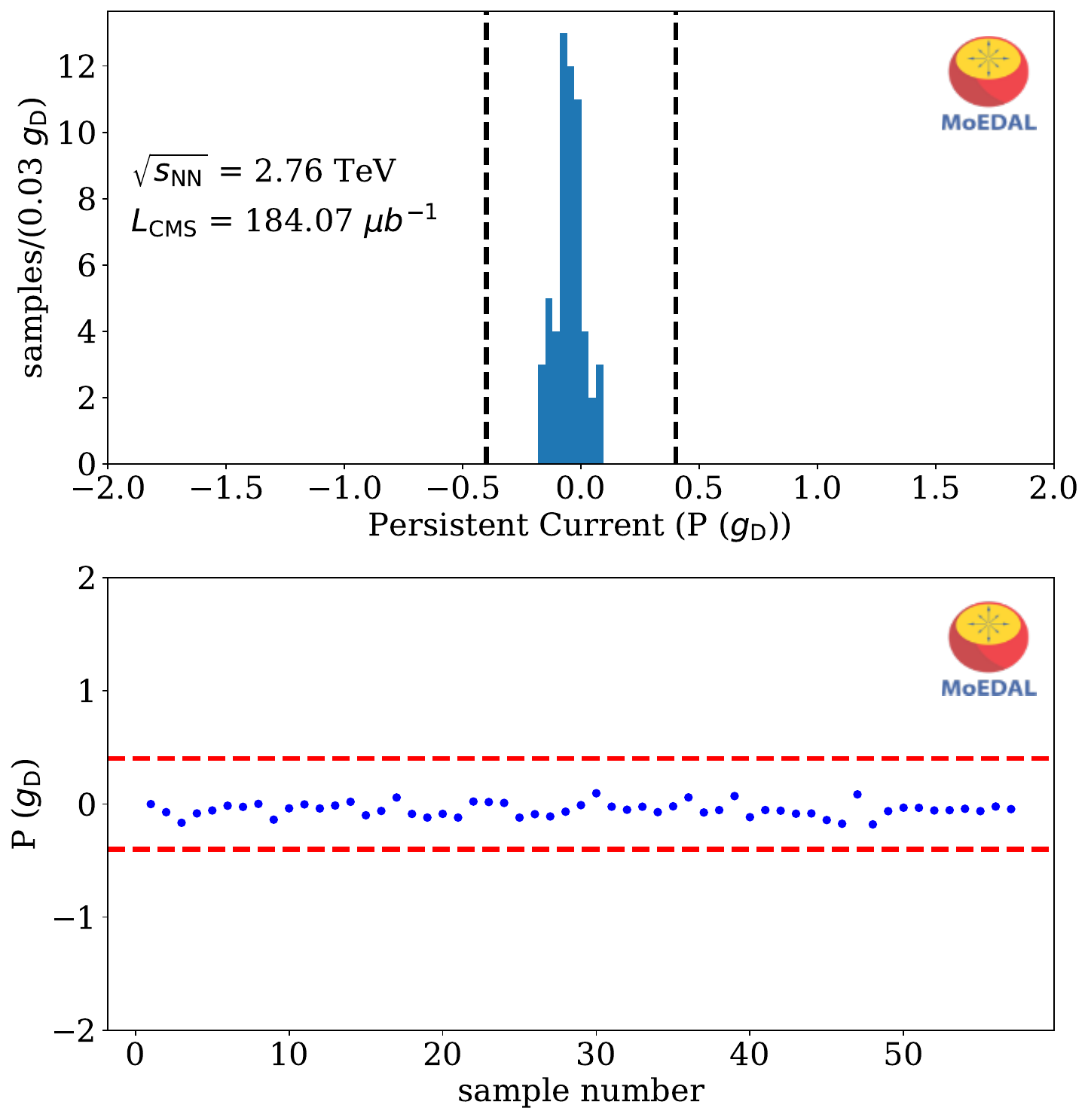}
\caption{(Top) Distribution of average persistent currents for the beam pipe samples. (Bottom) Average persistent current vs. sample number.} 
\label{fig:magnetometerHist}
\end{figure}

The beryllium section of the beam pipe was cut into 171 rings of 2-cm width each. The rings were further sub-divided into fragments. The fragments from three rings were then inserted into a plastic tube of dimension 2.54 cm $\times$ 2.54 cm $\times$ 19 cm and scanned together. The process was then repeated for fragments from another set of three rings. A total of 57 tubes with each set of fragments were passed through the SQUID multiple times (6--12 times) to scan for the presence of magnetic charges. Figure~\ref{fig:magnetometerHist} shows the results. No statistically significant signal was observed. To calibrate the device's response to magnetic charge, two independent methods were used, as described elsewhere~\cite{MoEDAL2016, moedal_prl2019}. The existence of an MM with $\mid g\mid$ $\geq$ 0.5 $g_{\mathrm{D}}$ in the trapping volume was excluded at more than 3$\sigma$.

\begin{figure*}[htpb]
\centering
    \includegraphics[width=0.95\textwidth]{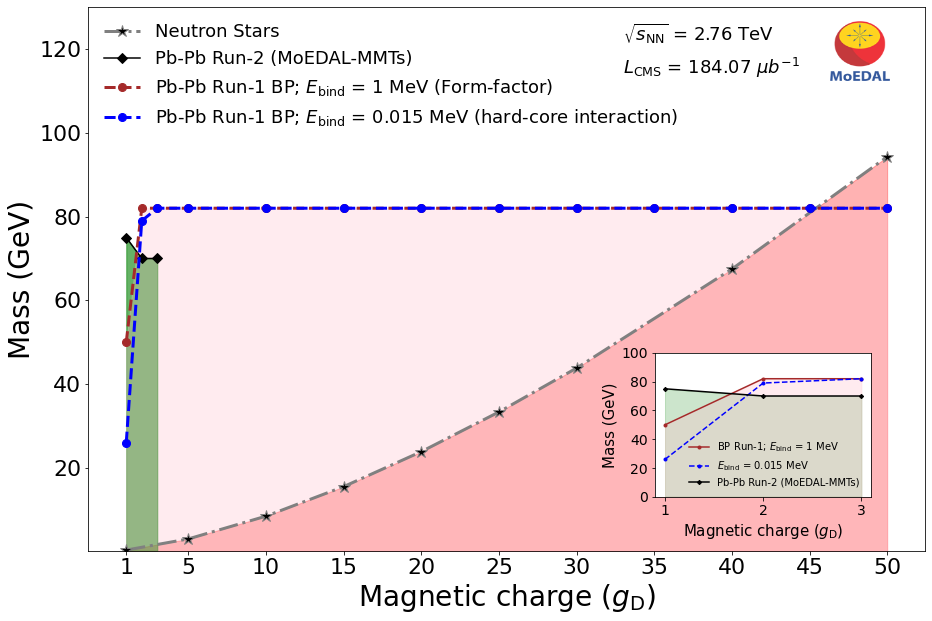}
\caption{The 95$\%$ C.L. exclusion region for MM search via the Schwinger effect in the CMS beam pipe (BP) exposed to Pb--Pb collisions during Run-1 of the LHC. The region shaded in green corresponds to the mass bounds from the MoEDAL search using its MMT detectors exposed to Pb--Pb collisions in Run-2~\cite{MoEDAL:2021vix}. The insert zooms in on the low-charge region. The limit from indirect searches for MMs produced by neutron stars~\cite{Gould:2021bre} is also shown, indicating that the current search provides the strongest available limits for charges up to 45 $g_{\mathrm{D}}$. The lines connecting the discrete charge values are to guide the eye.}
\label{fig:MassLimits}
\end{figure*}

Figure~\ref{fig:MassLimits} shows the 95\% confidence level (C.L.) exclusion region in the magnetic charge versus mass plane. The statistical significance of the limits is calculated from the Poisson statistics on $R_{\mathrm{exp}}$ and the uncertainties involved using an approach similar to that followed in Ref.~\cite{MoEDAL:2021vix}. The trapping efficiency increases with increasing magnetic charge up to 6 $g_{\mathrm{D}}$ and then starts decreasing. This is because high magnetic charges would be strongly accelerated by the CMS magnet, thus reducing the efficiency. However, the Schwinger LCFA cross section increases with increasing magnetic charge. The mass limits remain nearly constant because the mass of MMs that could be produced in Pb--Pb Run-1 collisions is limited by the energy in the electromagnetic field, calculated using Eq.~(\ref{eqn:maxField}). 

Our analysis combines the advantages of searching for MM production in heavy-ion collisions via the Schwinger effect with using a beam pipe as a trapping volume, which leads to unprecedented sensitivity to MMs with high magnetic charges. Our results are valid for both elementary and composite MMs. The limit from indirect searches for MMs produced by neutron stars~\cite{PRL_neutronStars} is also plotted in Fig.~\ref{fig:MassLimits}. The latter limits become stronger than the results of this work for magnetic charges above \mbox{45 $g_{\mathrm{D}}$}. 

In conclusion, the CMS beam pipe was exposed to 184.07 \textmu b$^{-1}$ Pb--Pb collisions at 2.76 TeV center-of-mass energy per collision in December 2011. The beam pipe acquired by MoEDAL from CMS was scanned for the presence of trapped magnetic charges using a SQUID magnetometer. No signal candidates were observed. The advantages of the ultraperipheral heavy-ion collisions and close proximity to the interaction point have enabled us to exclude, at 95\% C.L., the existence of composite or point-like MMs with masses up to 80 GeV, providing the strongest available constraint for magnetic charges from 2 to 45 $g_{\mathrm{D}}$.

\paragraph{Acknowledgments.} We thank CERN for the LHC operation, the CMS experiment for the beam pipe, as well as the support staff from our institutions without whom MoEDAL could not be operated. We acknowledge the invaluable assistance of the CMS collaboration. Computing support was provided by the GridPP Collaboration, in particular by the Queen Mary University of London and Liverpool grid sites. This work was supported by the UK Science and Technology Facilities Council, via the grants, ST/L000326/1, ST/L00044X/1, ST/N00101X/1, ST/P000258/1, ST/P000762/1, ST/T000732/1, ST/X000753/1 and ST/T000791/1; by the Generalitat Valenciana via the projects PROMETEO/2021/083, CIPROM/2021/054, CIPROM/2021/073, CIAPOS/2021/88; by Spanish MICIN Project PID2020-113334GB-I00/AEI/10.13039/501100011033, Project PID2021-122134NB-C21 funded by MCIN/AEI/10.13039/501100011033/ FEDER, UE and  MCIN/AEI/10.13039/501100011033, European Regional Development Fund Grant No. PID2019-105439 GB-C21; by MICIU via the mobility grant PRX22/00633; by the Physics Department of King's College London; by NSERC via a project grant; by the V-P Research of the University of Alberta (UofA); by the Provost of the UofA; by UEFISCDI (Romania); by the INFN (Italy); by the Estonian Research Council via a Mobilitas Pluss grant MOBTT5; by a Royal Society Dorothy Hodgkin Fellowship; and by the NSF grant 2309505 to the University of Alabama MoEDAL group. A. Rajantie was also supported by Institute for Particle Physics Phenomenology Associateship. I. Ostrovskiy thanks the Chinese Academy of Sciences (CAS) President's International Fellowship Initiative (PIFI) for the support.

\paragraph{Data Availability Statement.} The data supporting this study are available upon request. 

\bibliography{main}

\end{document}

%% file: commands.tex
\newcommand{\isot}[2]{$^{#2}$#1}
\newcommand{\isotbold}[2]{$^{\boldsymbol{#2}}$#1}
\newcommand{\xeiso}{\isot{Xe}{136}\xspace}
\newcommand{\thsrc}{\isot{Th}{228}\xspace}
\newcommand{\cosrc}{\isot{Co}{60}\xspace}
\newcommand{\rasrc}{\isot{Ra}{226}\xspace}
\newcommand{\cssrc}{\isot{Cs}{137}\xspace}
\newcommand{\betascale}  {$\beta$-scale}
\newcommand{\kevkgyr}  {keV$^{-1}$ kg$^{-1}$ yr$^{-1}$}
\newcommand{\nonubb}  {$0\nu \beta\!\beta$\xspace}
\newcommand{\nonubbbf}  {$\boldsymbol{0\nu \beta\!\beta}$\xspace}
\newcommand{\twonubb} {$2\nu \beta\!\beta$\xspace}
\newcommand{\bb} {$\beta\!\beta$\xspace}
\newcommand{\vadc} {ADC$_\text{V}$}
\newcommand{\uadc} {ADC$_\text{U}$}
\newcommand{\mus} {\textmu{}s}
\newcommand{\chisq} {$\chi^2$}
\newcommand{\mum} {\textmu{}m}
\newcommand{\red}[1]{{\xspace\color{red}#1}}
\newcommand{\blue}[1]{{\xspace\color{blue}#1}}
\newcommand{\RunTwoA}{Run 2a}
\newcommand{\RunTwo}{Run 2}
\newcommand{\RunTwoBC}{Runs 2b and 2c}
\newcommand{\SP}[1]{\textsuperscript{#1}}
\newcommand{\SB}[1]{\textsubscript{#1}}
\newcommand{\SPSB}[2]{\rlap{\textsuperscript{#1}}\SB{#2}}
\newcommand{\pmasy}[3]{#1\SPSB{$+$#2}{$-$#3}}
\newcommand{\matel}{$M^{2\nu}$}
\newcommand{\psfac}{$G^{2\nu}$}
\newcommand{\tbeta}{T$_{1/2}^{0\nu\beta\beta}$}
\newcommand{\exolimit}[1][true]{\pmasy{2.6}{1.8}{2.1}$ \cdot 10^{25}$}
\newcommand{\exomeasurement}{\tbeta{}= \exolimit{}~yr}
\newcommand{\U}{\text{U}}
\newcommand{\V}{\text{V}}
\newcommand{\X}{\text{X}}
\newcommand{\Y}{\text{Y}}
\newcommand{\Z}{\text{Z}}
\newcommand{\bqcm}{${\rm Bq~m}^{-3}$}
\newcommand{\nonunorm}{N_{{\rm Err, } 0\nu\beta\beta}}
\newcommand{\nonunum}{n_{0\nu\beta\beta}}
\newcommand{\cussim}[1]{$\sim$#1}
\newcommand{\halflife}[1]{$#1\cdot10^{25}$~yr}
\newcommand{\numspec}[3]{$N_{^{#2}\mathrm{#1}}=#3$}
\newcommand{\TD}[1]{\textcolor{red}{#1}}
\newcommand{\PI}{Phase~I\xspace}
\newcommand{\PII}{Phase~II\xspace}
\newcommand{\Rn}{radon\xspace}
\newcommand\Tstrut{\rule{0pt}{2.6ex}} 

%% file: moedal_authors_revtex.tex
\author{B.~Acharya}\thanks{Also at Int. Centre for Theoretical Physics, Trieste, Italy}\affiliation{Theoretical Particle Physics \& Cosmology Group, Physics Dept., King’s College London, UK}
\author{J.~Alexandre}\affiliation{Theoretical Particle Physics \& Cosmology Group, Physics Dept., King’s College London, UK}
\author{S.C.~Behera}\affiliation{Department of Physics and Astronomy, University of Alabama, Tuscaloosa, Alabama, USA}
\author{P.~Benes}\affiliation{IEAP, Czech Technical University in Prague, Czech Republic}
\author{B.~Bergmann}\affiliation{IEAP, Czech Technical University in Prague, Czech Republic}
\author{S.~Bertolucci}\affiliation{INFN, Section of Bologna, Bologna, Italy}
\author{A.~Bevan}\affiliation{School of Physics and Astronomy, Queen Mary University of London, UK}
\author{R.~Brancaccio}\affiliation{INFN, Section of Bologna \& Department of Physics \& Astronomy, University of Bologna, Italy}
\author{H.~Branzas}\affiliation{Institute of Space Science, Bucharest - Magurele, Romania}
\author{P.~Burian}\affiliation{IEAP, Czech Technical University in Prague, Czech Republic}
\author{M.~Campbell}\affiliation{Experimental Physics Department, CERN, Geneva, Switzerland}
\author{S.~Cecchini}\affiliation{INFN, Section of Bologna, Bologna, Italy}
\author{Y.~M.~Cho}\affiliation{Center for Quantum Spacetime, Sogang University, Seoul, Korea}
\author{M.~de~Montigny}\affiliation{Physics Department, University of Alberta, Edmonton, Alberta, Canada}
\author{A.~De~Roeck}\affiliation{Experimental Physics Department, CERN, Geneva, Switzerland}
\author{J.~R.~Ellis}
\affiliation{Theoretical Particle Physics \& Cosmology Group, Physics Dept., King’s College London, UK}\affiliation{Theoretical Physics Department, CERN, Geneva, Switzerland}
\author{M.~Fairbairn}\affiliation{Theoretical Particle Physics \& Cosmology Group, Physics Dept., King’s College London, UK}
\author{D.~Felea}\affiliation{Institute of Space Science, Bucharest - Magurele, Romania}
\author{M.~Frank}\affiliation{Department of Physics, Concordia University, Montreal, Quebec, Canada}
\author{O.~Gould}\affiliation{University of Nottingham, Nottingham, UK}
\author{J.~Hays}\affiliation{School of Physics and Astronomy, Queen Mary University of London, UK}
\author{A.M.~Hirt}\affiliation{Department of Earth Sciences, Swiss Federal Institute of Technology, Zurich, Switzerland}
\author{D.~L.-J.~Ho}\affiliation{Department of Physics, Imperial College London, UK}
\author{P.~Q.~Hung}\affiliation{Department of Physics, University of Virginia, Charlottesville, VA, USA}
\author{J.~Janecek}\affiliation{IEAP, Czech Technical University in Prague, Czech Republic}
\author{M.~Kalliokoski}\affiliation{Helsinki Institute of Physics, University of Helsinki, Helsinki, Finland}
\author{D.~H.~Lacarr\`{e}re}\affiliation{Experimental Physics Department, CERN, Geneva, Switzerland}
\author{C.~Leroy}\affiliation{Departement de Physique, Universite de Montreal, Quebec, Canada}
\author{G.~Levi}\affiliation{INFN, Section of Bologna \& Department of Physics \& Astronomy, University of Bologna, Italy}
\author{A.~Margiotta}\affiliation{INFN, Section of Bologna \& Department of Physics \& Astronomy, University of Bologna, Italy}
\author{R.~Maselek}\affiliation{Institute of Theoretical Physics, University of Warsaw, Warsaw, Poland}
\author{A.~Maulik}\affiliation{INFN, Section of Bologna, Bologna, Italy}\affiliation{Physics Department, University of Alberta, Edmonton, Alberta, Canada}
\author{N. ~Mauri}\affiliation{INFN, Section of Bologna \& Department of Physics \& Astronomy, University of Bologna, Italy}
\author{N. E. ~Mavromatos}\thanks{Also at Department of Physics, School of Applied Mathematical and Physical Sciences, National Technical University of Athens, Athens, Greece}\affiliation{Theoretical Particle Physics \& Cosmology Group, Physics Dept., King’s College London, UK}
\author{L.~Millward}\affiliation{School of Physics and Astronomy, Queen Mary University of London, UK}
\author{V.~A.~Mitsou}\thanks{Also at Department of Physics, School of Applied Mathematical and Physical Sciences, National Technical University of Athens, Athens, Greece}\affiliation{IFIC, Universitat de Valencia - CSIC, Valencia, Spain}
\author{E.~Musumeci}\affiliation{IFIC, Universitat de Valencia - CSIC, Valencia, Spain}
\author{I.~Ostrovskiy}\thanks{Currently at Institute of High Energy Physics, Beijing, China}\email[Corresponding author: ]{iostrovskiy@ua.edu}\affiliation{Department of Physics and Astronomy, University of Alabama, Tuscaloosa, Alabama, USA}
\author{P.-P.~Ouimet}\affiliation{Physics Department, University of Regina, Regina, Saskatchewan, Canada}
\author{J.~Papavassiliou}\affiliation{IFIC, Universitat de Valencia - CSIC, Valencia, Spain}
\author{L.~Patrizii}\affiliation{INFN, Section of Bologna, Bologna, Italy}
\author{G.~E.~P\u{a}v\u{a}la\c{s}}\affiliation{Institute of Space Science, Bucharest - Magurele, Romania}
\author{J.~L.~Pinfold}\affiliation{Physics Department, University of Alberta, Edmonton, Alberta, Canada}
\author{L.~A.~Popa}\affiliation{Institute of Space Science, Bucharest - Magurele, Romania}
\author{V.~Popa}\affiliation{Institute of Space Science, Bucharest - Magurele, Romania}
\author{M.~Pozzato}\affiliation{INFN, Section of Bologna, Bologna, Italy}
\author{S.~Pospisil}\affiliation{IEAP, Czech Technical University in Prague, Czech Republic}
\author{A.~Rajantie}\affiliation{Department of Physics, Imperial College London, UK}
\author{R.~Ruiz~de~Austri}\affiliation{IFIC, Universitat de Valencia - CSIC, Valencia, Spain}
\author{Z.~Sahnoun}\affiliation{INFN, Section of Bologna \& Department of Physics \& Astronomy, University of Bologna, Italy}
\author{M.~Sakellariadou}\affiliation{Theoretical Particle Physics \& Cosmology Group, Physics Dept., King’s College London, UK}
\author{K.~Sakurai}\affiliation{Institute of Theoretical Physics, University of Warsaw, Warsaw, Poland}
\author{S.~Sarkar}\affiliation{Theoretical Particle Physics \& Cosmology Group, Physics Dept., King’s College London, UK}
\author{G.~Semenoff}\affiliation{Department of Physics, University of British Columbia, Vancouver, British Columbia, Canada}
\author{A.~Shaa}\affiliation{Physics Department, University of Alberta, Edmonton, Alberta, Canada}
\author{G.~Sirri}\affiliation{INFN, Section of Bologna, Bologna, Italy}
\author{K.~Sliwa}\affiliation{Department of Physics and Astronomy, Tufts University, Medford, Massachusetts, USA}
\author{R.~Soluk}\affiliation{Physics Department, University of Alberta, Edmonton, Alberta, Canada}
\author{M.~Spurio}\affiliation{INFN, Section of Bologna \& Department of Physics \& Astronomy, University of Bologna, Italy}
\author{M.~Staelens}\affiliation{IFIC, Universitat de Valencia - CSIC, Valencia, Spain}
\author{M.~Suk}\affiliation{IEAP, Czech Technical University in Prague, Czech Republic}
\author{M.~Tenti}\affiliation{INFN, Section of Bologna, Bologna, Italy}
\author{V.~Togo}\affiliation{INFN, Section of Bologna, Bologna, Italy}
\author{J.~A.~Tuszy\'{n}ski}\affiliation{Physics Department, University of Alberta, Edmonton, Alberta, Canada}
\author{A.~Upreti}\affiliation{Department of Physics and Astronomy, University of Alabama, Tuscaloosa, Alabama, USA}
\author{V.~Vento}\affiliation{IFIC, Universitat de Valencia - CSIC, Valencia, Spain}
\author{O.~Vives}\affiliation{IFIC, Universitat de Valencia - CSIC, Valencia, Spain}